\documentclass[a4paper,10pt]{article}
\usepackage{graphicx}
\usepackage{fancyhdr}
\usepackage{xcolor}
\usepackage{amsmath}
\usepackage{multicol}
\usepackage{color}

\begin{document}
\noindent
{\it 11th Hel.A.S Conference}\\
\noindent
{\it Athens, 8-12 September, 2013}\\
\noindent
%
%
\noindent
\underline{~~~~~~~~~~~~~~~~~~~~~~~~~~~~~~~~
~~~~~~~~}
\vskip 0.9cm
%
%
\begin{center}
{\Large\bf
Multi-frequency linear and circular radio polarization monitoring of jet emission elements in $Fermi$ blazars
}
\vskip 0.4cm
%
%
{\it
I.\,Myserlis$^1$, E.\,Angelakis$^1$, L.\,Fuhrmann$^1$, V.\,Pavlidou$^{2,3}$, I.\,Nestoras$^1$, V.\,Karamanavis$^1$, A.\,Kraus$^1$, J.\,A.\,Zensus$^1$ 
}\\
%
%
\begin{small}
$^1$ Max-Planck-Institut f\"uˆr Radioastronomie, Auf dem H\"ugel 69, 53121 Bonn, Germany \\
$^2$ Foundation for Research and Technology - Hellas, IESL, Voutes, 7110 Heraklion, Greece \\
$^3$ Department of Physics and Institute for Plasma Physics, University of Crete, 71003, Heraklion, Greece 
\end{small}

\end{center}
\vskip 0.5cm
%
%
\noindent
{\bf Abstract: }
Radio emission in blazars -- the aligned subset of Active Galactic Nuclei (AGN) -- is produced
by synchrotron electrons moving relativistically in their jet's magnetic field.
Under the
assumption of some degree of uniformity of the field, the emission can be highly polarized
-- linearly and circularly. In the radio regime, the observed variability is in most of the cases
attributed to flaring events undergoing opacity evolution, i.e. transitions from optically thick to thin
emission (or vice versa). These transistions have a specific signature in the polarization parameter space (angle and magnitude) which can be
traced with high cadence polarization monitoring and provide us with a unique probe of the microphysics
of the emitting region. Here we present the full Stokes analysis of radio emission from blazars observed in the
framework of the F-GAMMA program and discuss the case study of PKS\,1510$-$089 which has shown a prominent
polarization event around MJD 55900.


\section{Introduction}
\label{sec: intro}

Blazars -- including Flat Spectrum Radio Quasars (FSRQs) and BL Lac objects --
comprise the most extreme manifestation of the AGN phenomenon.  The combination of
relativistic speeds and small angles of their jet to our line of sight (a few degrees \cite{blandford1979}),
results in an extreme phenomenology. Their emission is beamed
and boosted showing intense variability at all time scales (from years down to hours), very high 
brightness temperatures, highly superluminal apparent speeds and
others. They emit a remarkably broadband spectral energy distribution (SED) spanning from
radio to TeV energies,  
 with variability seen throughout its entirety. The broadness of their SED is caused by
incoherent synchrotron processes involving most likely electrons which -- at second order
-- are Inverse-Compton upscattering photons to very high energies. The synchrotron
part is intrinsically polarized to a degree that depends on opacity \cite{pacholczyk1977}.

The radio tail of the synchrotron component is practically entirely attributed to the
large scale jet. The variability seen in these frequencies in most of the cases shows
signature of spectral evolution and transition from optically thick to thin regime. This
alone makes the radio emission, and especially its polarized part, an excellent probe 
of the microphysics of extragalactic jets. In the following, and after a brief review of the 
mechanisms believed to be responsible for polarization, we show a case study to demonstrate 
the insight that such studies can provide and discuss the techniques and methods that are 
involved in this endeavor.

\section{Polarized emission from relativistic jets}
\label{sec: theory}
The AGN radio emission is attributed to incoherent synchrotron radiation of relativistic
electrons and therefore, it is expected to be polarized. Under certain conditions, a rather
high degree of linear 
or circular 
polarization could be observed. The condition
for significant polarization is the degree of uniformity of the magnetic field implying that 
polarization probes uniformly magnetized regions of the plasma and trace their behavior in time.
Multi-band linear and circular radio polarization monitoring is then an
invaluable tool for the investigation of the physical properties of AGN jets such as the 
topology and magnitude of their magnetic fields \cite{saikia1988}, their composition \cite{jones1977a,jones1977b}
and structural characteristics of their environment \cite{antonucci1985}.

The degree of linear polarization in the optically thin case and under the assumption of
uniform magnetic field, is expected to be as high as almost 72\,\% ($m^\mathrm{(thin)}_\mathrm{l} =
(|\alpha|+1)/(|\alpha|+\frac{5}{3})$ with $\alpha$ the spectral index defined as $S_\nu
\propto \nu^{\alpha}$ \cite{pacholczyk1977}). The electric vector (EV) is perpendicular to
the projection of the magnetic field onto the plane of the sky. The emission is expected
to be also circularly polarized with its handedness to depend on the viewing angle of the
system. The expected circular polarization degree is generally a tiny fraction of the
linear polarization 
and has a frequency dependence of the form:
$m^\mathrm{(thin)}_\mathrm{c} \propto \nu^{-\frac{1}{2}}$.

At low radio frequencies -- where the synchrotron emission is self-absorbed -- the degree
of linear polarization drops to $m^\mathrm{(thick)}_\mathrm{l} = 3/(12|\alpha| + 19) \approx 11\,\%$
with the EV parallel to the projection of the magnetic field onto the plane of the sky,
and the circular polarization degree follows the same spectral dependence though with the opposite 
handedness \cite{pacholczyk1977}.
In the radio regime the spectrum is expected to undergo the transition between
optically thin and thick emission \cite{tuerler2000}. This makes multi-frequency
polarization observations a unique
diagnostic tool for the detailed study of the radiative mechanism.

High-cadence polarization monitoring, can trace the dynamics of the associated physical
characteristics \cite{allers1989}.
Polarization variability takes place mainly in two domains: (a)
the polarization amplitude and (b) the electric vector position angle (EVPA). EVPA variability
has been observed in the past in the form of polarization angle swings 
(e.g. \cite{aller1981,marcher2008,abdo2010}). 
Such swings can be attributed to either of two mechanisms:
(a) \textbf{Geometrical swing:} The swing is caused by the motion of a synchrotron
  emitting cloud of electrons in a helical path and the rotation can have theoretically an
  arbitrary magnitude depending on the helical path and the line of sight.
(b) \textbf{Radiative ``swing'':} The rotation is caused by the transition from the
  optically thick to thin regime (or vice versa) of the synchrotron self-absorbed spectrum.
  As a jet emission element expands, its density and consequently its optical thickness, $\tau$, decrease. 
  When $\tau\approx1$, the linear as well as the circular polarization degrees drop to zero before gaining
  the values associated with the optically thin (or thick) regime mentioned earlier, while the EVPA changes by exactly $90^\circ$. 
  
The described mechanisms can also operate simultaneously. However, it is essential that
only in the radio regime where dynamic outbursting activity induces opacity evolution --
can radiative swings be studied. 

\section{The F-GAMMA program and radio polarization data analysis}
 \label{sec: fgamma}

 Despite the unique potential of the polarimetric studies and especially of polarization
 monitoring investigations, the observational challenges involved are often
 insurmountable. The F-GAMMA program is a radio multi-frequency {\it Fermi} blazar monitoring
 program operating since January 2007 \cite{angelakis2008,fuhrmann2007}, providing a database
 suitable for such studies. Observations are conducted with 3 main facilities: the 100-m
 Effelsberg, the 30-m IRAM and the 12-m APEX telescopes. The datasets consist of light
 curves at 11 radio frequencies from 2.64 to 345 GHz, linear polarization for 5 frequencies
 between 2.64 and 14.6 GHz and circular polarization for 7 frequencies between 2.64 to
 43 GHz for $\sim$\,60 $\gamma$-ray blazars with a mean cadence of 1.1 months.

 The difficulty of polarization measurements of AGN is apparent due to the low degree of
 polarization in the radio band. As an example, at 5\,GHz, 75\,\% of the F-GAMMA sources are
 linearly polarized with median linear polarization degree ($\widetilde{m}_\mathrm{l}$) of 3.1\,\% and 2\,\% are
 circularly polarized with median circular polarization degree ($\widetilde{m}_\mathrm{c}$) of 0.4\,\%. At 10\,GHz, 39\,\% of
 the sample is linearly polarized with $\widetilde{m}_\mathrm{l} = 3.6\,\%$ and 5\,\%
 is circularly polarized with $\widetilde{m}_\mathrm{l} = 0.4\,\%$\footnote{These values
 refer to $3\sigma$ detections of the linear and circular polarization flux.}.
 
 The state of polarization can be described by the Stokes vector, $\mathbf{S} =
 \left(I,Q,U,V\right)$, where $I$ is the total intensity, $Q$ and $U$ describe the linear
 and $V$ the circular polarization characteristics. Instrumental effects which alter the
 polarization state of the incident radiation, can be described by the M\"uller
 matrix, $\bf{M}$, a transfer function between the real and the observed
 Stokes parameters: $\mathbf{S_{obs}} = \mathbf{M} \cdot \mathbf{S_{real}}$.

The applied analysis to correct for instrumental effects can be broken down to: (a) observe sources
of known polarization parameters (standards), (b) with knowns the observed and the expected Stokes vectors, compute the
elements of the M\"uller matrix. For an over-determined system, a least-squares fit
method is applied and (c) apply the inverse of the computed M\"uller matrix to the observed
Stokes vector for the target sources \cite{turlo1985}.

The reduction scheme described above is usually applied only
for linear polarization measurements owing to the lack of circular polarization
standards. The advantage however in our case is the fact that the receivers are equipped
with circularly polarized feeds which allow satisfactory measurements after precise
gain correction using observations of unpolarized sources. A number of
sources with significant circular polarization flux have been detected as well as sources with stable
circular polarization parameters which could be used as standards, such as 3C295
and 3C48 which have circular polarization degrees of $0.6\pm0.1\,\%$ and $0.6\pm0.2\,\%$ at 5\,GHz respectively.

\section{PKS\,1510$-$089: a case study}
\begin{figure}[h!]
\centering
\includegraphics[width=0.7\textwidth]{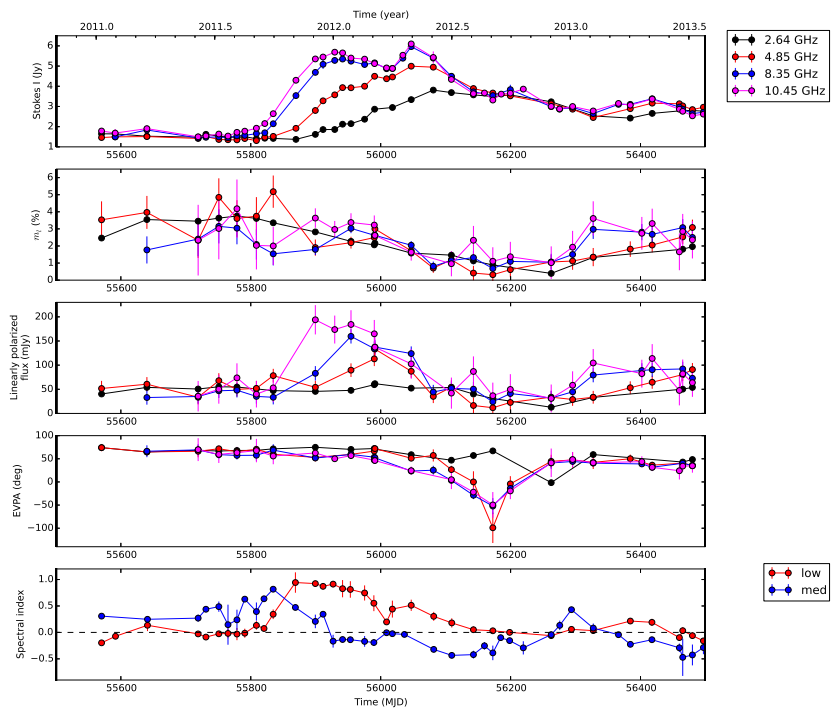}
\caption{\footnotesize{Polarization parameters of the blazar PKS\,1510$-$089, as obtained
    with the F-GAMMA program.  From top to bottom: Stokes $I$, linear polarization degree
    $m_\mathrm{l}$, linearly polarized flux, EVPA and spectral index in two radio bands
    ($\alpha_{2.6}^{8.4}$ and $\alpha_{10.5}^{23.1}$).}}
\label{1510}
\end{figure}
PKS\,1510$-$089 is among the best studied blazars (FSRQ). It has shown geometrically
induced rotations of the EVPA \cite{marscher2010}. As an example case study, we discuss
the radio polarimetric data that display a very interesting behavior.  In
Fig.\,\ref{1510}, from top to bottom, is shown: Stokes $I$, the degree of linear
polarization $m_\mathrm{l}$, the linearly polarized flux, the EVPA and the spectral index in two
radio bands: low ($\alpha_{2.6}^{8.4}$) and intermediate ($\alpha_{10.5}^{23.1}$).

What is immediately visible in the Stokes $I$ panel is that the source underwent a major
flaring event from MJD 55800 to MJD 56200, less discernible towards lower frequencies.
The highest peak appears around MJD 56050 and with a phase difference between the 4
frequencies, due to opacity effects. The spectrum shows profound evolution during the
flare. In the low band, it starts off as flat ($\alpha\approx0$) and hardens (becoming
optically thicker) during the outburst before flattening again.
A major $\gamma$-ray event preceded the radio one for about 140 days at $\mathrm{MJD}\approx55860$ \cite{orienti2013}.

The polarization follows this evolution closely. During the flaring event, the
polarization degree remains fairly stable ($\approx 2-3$\,\%) and the polarized flux shares
the characteristics of Stokes $I$. The stability of the polarization degree over the
flaring event implies that no opacity transition takes place as it can also be seen in the
spectral index behavior which remains optically thick ($\alpha > 0$).
The EVPA shows a very interesting behavior. At the beginning of 2011 ($\mathrm{MJD}\approx55600$) it is $\sim65^\circ$,
roughly perpendicular to the jet axis \cite{marscher2010}. Around MJD 56170 it shows two
prominent rotations.
During the former rotation (up to MJD 56170), the EVPA rotates for $\approx125^\circ$ ($65^\circ$ to
$-60^\circ$) in total with a pace of $1.2^\circ$ per day at 5\,GHz and $0.5^\circ$ per day
at 10\,GHz. During the latter one (from MJD 56170 on) it rotates in the opposite
sense until it reaches $\approx40^\circ$ ($-60^\circ$ to $40^\circ$) with a pace of $1.2^\circ$
per day at 5\,GHz and $0.8^\circ$ per day at 10\,GHz.

In order to explain the observed behavior, we 
assume that an optically thick emission element which is 
responsible for the major Stokes $I$ event undergoes a slow geometrically induced rotation (Sec.\,\ref{sec: theory}),
from $65^\circ$ to $40^\circ$, first at 10\,GHz (up to $\mathrm{MJD}\approx56000$) and then at 5\,GHz
(up to $\mathrm{MJD}\approx56100$), because the polarization degree remains stable and the spectral index
remains thick all the time.  Then the same emission element becomes optically thin and EVPA rotates by
$90^\circ$ becoming roughly parallel to the jet axis \cite{marscher2010} (fast
rotation before MJD 56170, $40^\circ$ to $-50^\circ$ at 10\,GHz) during a concurrent minimization of
the polarization degree and flux, a characteristic of radiative swings as described in
Sec.\,\ref{sec: theory}.  Finally, a subsequent optically thick emission element, possibly
the one responsible for the short flare around MJD 56420,
rotates the EVPA again by $90^\circ$, perpendicular to the jet axis \cite{marscher2010} (fast
rotation after MJD 56170, $-50^\circ$ to $40^\circ$ at 10\,GHz).  Both fast rotations around MJD 56170 are then of the radiative
kind (Sec.\,\ref{sec: theory}) as it is also suggested by the fact that their paces are
very similar.

Assuming that the radiative swing does not occur simultaneously
with the optical depth transition, as it is derived in \cite{pacholczyk1977}, an alternative interpretation would be 
that the first emission element has already the optically thin polarization characteristics at the beginning of our measurements.
This could explain the high polarization degree
. Then the second 
emission element would be responsible for both radiative EVPA swings around MJD 56170. The second element which is initially
optically thick, rotates the EVPA to $-50^\circ$ ($90^\circ$ rotation) and reduces the polarization degree. Later it 
becomes optically thin, rotating the EVPA back to $40^\circ$ and increasing the polarization degree once more.

\section{Summary}

The prominent variability and the associated spectral evolution of blazars 
carry a characteristic signature in their polarized radio emission, providing a unique probe of the microphysics
of the emitting region. After an introduction
for the study of blazar physics using multi-band high-cadence polarization monitoring, we
described the full Stokes analysis of the F-GAMMA database.
Finally, we focused on the case study of PKS\,1510$-$089 which showed an
interesting polarization event around MJD 55900.

\noindent

{\bf Acknowledgements:} I.M. is supported for this research through a stipend from the 
International Max Planck Research School (IMPRS) for Astronomy and Astrophysics at the Universities
of Bonn and Cologne. Based on observations with the 100-m telescope of the MPIfR
(Max-Planck-Institut f\"ur Radioastronomie) at Effelsberg. The authors thank the
internal referee at the MPIfR, Dr. Andrei Lobanov, for his constructive comments.

\end{document}